\documentclass[twocolumn]{aastex63}

\usepackage[T1]{fontenc}

\newcommand{\package}[1]{\textsl{#1}}
\newcommand{\gaia}{\textsl{Gaia}}
\newcommand{\hst}{\textsl{HST}}
\newcommand{\pans}{\textsl{Pan-STARRS}}

\newcommand{\kms}{\ensuremath{\textrm{km}\,\textrm{s}^{-1}}}
\newcommand{\masyr}{\ensuremath{\textrm{mas}\,\textrm{yr}^{-1}}}
\newcommand{\feh}{\ensuremath{\textrm{[Fe/H]}}}
\newcommand{\afe}{\ensuremath{\textrm{[$\alpha$/Fe]}}}

\shorttitle{a comoving spur of gd-1}
\shortauthors{bonaca et al.}

\usepackage{amsmath}

\begin{document}\sloppy\sloppypar\raggedbottom\frenchspacing 

\title{High-resolution spectroscopy of the GD-1 stellar stream\\localizes the perturber near the orbital plane of Sagittarius}

\correspondingauthor{Ana~Bonaca}
\email{ana.bonaca@cfa.harvard.edu}

\author[0000-0002-7846-9787]{Ana~Bonaca}
\affil{Center for Astrophysics | Harvard \& Smithsonian, 60 Garden Street, Cambridge, MA 02138, USA}

\author[0000-0002-1590-8551]{Charlie~Conroy}
\affil{Center for Astrophysics | Harvard \& Smithsonian, 60 Garden Street, Cambridge, MA 02138, USA}

\author[0000-0003-2866-9403]{David~W.~Hogg}
\affiliation{Center for Cosmology and Particle Physics, Department of Physics, New York University}
\affiliation{Center for Data Science, New York University}
\affiliation{Max-Planck-Institut f\"ur Astronomie, Heidelberg}
\affiliation{Center for Computational Astrophysics, Flatiron Institute, 162 Fifth Avenue, NY 10010, USA}

\author{Phillip~A.~Cargile}
\affil{Center for Astrophysics | Harvard \& Smithsonian, 60 Garden Street, Cambridge, MA 02138, USA}

\author{Nelson~Caldwell}
\affil{Center for Astrophysics | Harvard \& Smithsonian, 60 Garden Street, Cambridge, MA 02138, USA}

\author[0000-0003-3997-5705]{Rohan~P.~Naidu}
\affil{Center for Astrophysics | Harvard \& Smithsonian, 60 Garden Street, Cambridge, MA 02138, USA}

\author[0000-0003-0872-7098]{Adrian~M.~Price-Whelan}
\affiliation{Center for Computational Astrophysics, Flatiron Institute, 162 Fifth Avenue, NY 10010, USA}

\author[0000-0002-5065-9896]{Joshua~S.~Speagle}
\affil{Center for Astrophysics | Harvard \& Smithsonian, 60 Garden Street, Cambridge, MA 02138, USA}

\author{Benjamin~D.~Johnson}
\affil{Center for Astrophysics | Harvard \& Smithsonian, 60 Garden Street, Cambridge, MA 02138, USA}

\begin{abstract}\noindent 
The $100^\circ$-long thin stellar stream in the Milky Way halo, GD-1, has an ensemble of features that may be due to dynamical interactions.
Using high-resolution MMT/Hectochelle spectroscopy we show that a spur of GD-1-like stars outside of the main stream are kinematically and chemically consistent with the main stream.
In the spur, as in the main stream, GD-1 has a low intrinsic radial velocity dispersion, $\sigma_{V_r}\lesssim1\,\kms$, is metal-poor, $\feh\approx-2.3$, with little \feh\ spread and some variation in \afe\ abundances, which point to a common globular cluster progenitor.
At a fixed location along the stream, the median radial velocity offset between the spur and the main stream is smaller than $0.5\,\kms$, comparable to the measurement uncertainty.
A flyby of a massive, compact object can change orbits of stars in a stellar stream and produce features like the spur observed in GD-1.
In this scenario, the radial velocity of the GD-1 spur relative to the stream constrains the orbit of the perturber and its current on-sky position to $\approx5,000\,\deg^2$.
The family of acceptable perturber orbits overlaps the stellar and dark-matter debris of the Sagittarius dwarf galaxy in present-day position and velocity.
This suggests that GD-1 may have been perturbed by a globular cluster or an extremely compact dark-matter subhalo formerly associated with Sagittarius.
\end{abstract}

\keywords{%
stars:~kinematics~and~dynamics
  ---
Galaxy:~halo
  ---
Galaxy:~kinematics~and~dynamics
}

\section{Introduction}
\label{sec:intro}

The preeminent cosmological model predicts that galaxies like the Milky Way contain a myriad of non-luminous clumps of dark matter \citep[e.g.,][]{diemand2008, springel2008}.
Masses of these dark-matter subhalos are $\gtrsim4$ orders of magnitude lower than the total mass of the Milky Way, so they are expected to have a negligible effect on most stars in the Galaxy \citep[e.g.,][]{hopkins2008, donghia2010}.
However, even low-mass subhalos would leave evidence of interaction with stellar streams, the tidal debris of luminous satellites.
Numerical experiments have shown that subhalo encounters can heat up streams \citep[e.g.,][]{johnston2002, ibata2002}, produce gaps in their density profiles \citep[e.g.,][]{sgv2008,yoon2011}, and cause stream folds \citep[e.g.,][]{carlberg2009}.

Until recently, observations of stellar streams in the Milky Way were insufficient to allow robust searches for signatures of dark-matter subhalos \citep[cf.][]{carlberg2012, ibata2016}.
Now, proper motions from the \gaia\ mission \citep{gdr2} have revolutionized our ability to discover \citep[e.g.,][]{malhan2018,meingast2019} and characterize stellar streams \citep[e.g.,][]{bonaca2019b,shipp2019}.
Using \gaia\ data, \citet{pwb} studied a nearby, retrograde stellar stream GD-1 \citep{grillmair2006}, produced the cleanest map of a stream in the Milky Way and confidently identified several underdensities, as well as stars outside of the main stream \citep[see also][]{malhan2019b, deboer2019}.
\citet{bonaca2019a} created dynamical models of GD-1 that, following an encounter with a massive object, form a stream gap and an adjacent spur of stars that  quantitatively match the observed features.
With no known luminous object having approached GD-1 sufficiently close, there is a possibility that GD-1 was perturbed by a dark-matter subhalo.

\begin{figure*}
\begin{center}
\includegraphics[width=0.99\textwidth]{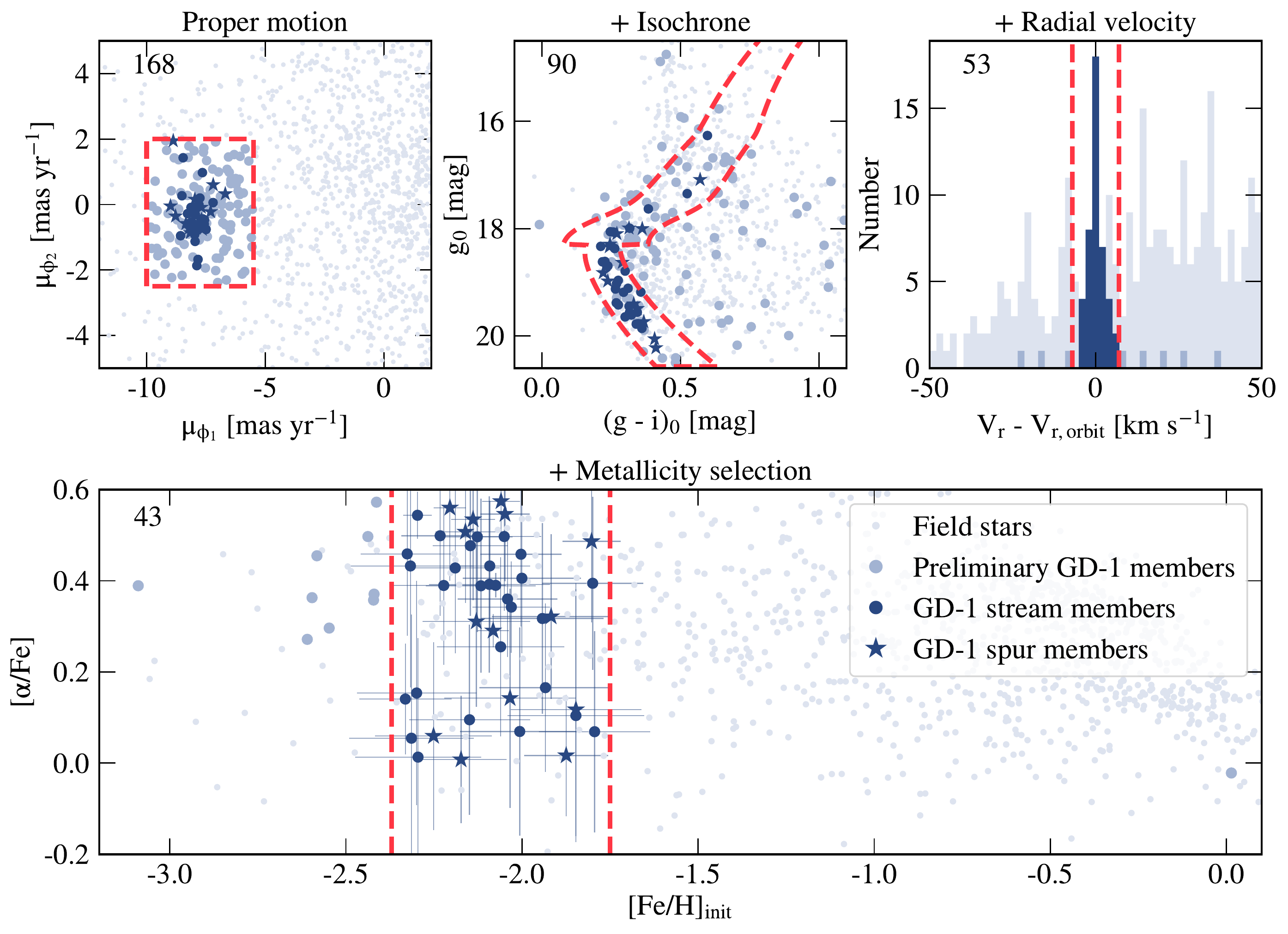}
\end{center}
\caption{We defined membership to the GD-1 stream with four selection criteria: (1) proper motion box (top left), (2) isochrone box (top middle), (3) small radial velocity offset from the GD-1 orbit (top right), (4) low metallicity (bottom).
Starting clockwise with the proper motion selection, panels add selections marked with pink dashed lines and decrease membership to the number in the top right of the panel.
In each panel, the non-members, preliminary and high-probability members are shown in light, medium and dark blue, respectively.
The GD-1 spur (stars) is kinematically and chemically consistent with the main stream (circles).
}
\label{fig:members}
\end{figure*}

Precise kinematic data are required to test whether the spur-and-gap feature in GD-1 was indeed formed in an interaction with a massive, dark object \citep{bonaca2019a}.
Until now, radial velocities have only been available in the main GD-1 stream and at low precision \citep{koposov2010,huang2019}.
In Section~\ref{sec:spec} we present the high-resolution spectroscopy from MMT/Hectochelle, which we used to define a sample of highly probable GD-1 members in the main stream and in the spur (\S\ref{sec:membership}).
These data show that the spur is kinematically aligned with the GD-1 stream (\S\ref{sec:kinematics}).
The small relative velocity between the stream and the spur can be explained within the impact scenario, but only if a perturber is on a specific set of orbits (\S\ref{sec:discussion}), which improves prospects of locating dark objects within the Milky Way purely from their interactions with stellar streams.

\section{Spectroscopy}
\label{sec:spec}

We observed the GD-1 stellar stream using the MMT/Hectochelle multi-object spectrograph \citep{szentgyorgyi2011}.
Focusing on the perturbed area at $\phi_1\approx-40^\circ$ \citep[$\phi_{1,2}$ are coordinates oriented along and perpendicular to GD-1, respectively;][]{koposov2010}, we targeted 4 fields in the main stream, and 4 fields in the lower-density spur (Figure~\ref{fig:vr}, top).
Using the \gaia--\pans\ cross-matched catalog, we selected retrograde stars as science targets, first prioritizing stars on the GD-1 main sequence, and then its red giant branch \citep[see][]{pwb}.
On average, we dedicated $\gtrsim170$ fibers to science targets per field, for a total of 1409 science spectra.
Up to 40 of the remaining fibers were used to estimate the sky emission.
We used the \texttt{RV31} filter covering the Mg~b triplet and observed each field for 2.25~hours (except for stream field at $\phi_1\approx-34^\circ$ which was observed for 2~hours due to scheduling constraints).
With $2\times3$ spatial and spectral binning of the CCD pixels, we achieved a signal-to-noise ratio $S/N\approx2$ at $g=20$ and an effective resolution $R\approx32,000$.

The 2D spectra were reduced by \package{HSRED}~v2.1\footnote{\url{https://bitbucket.org/saotdc/hsred/}}.
This pipeline flat-fields, wavelength-calibrates with respect to ThAr lamp spectra, extracts 1D spectra and subtracts the sky emission.
We then used the \package{MINESweeper} code \citep{cargile2019} to forward-model the processed 1D spectra and infer stellar parameters, including radial velocities, [Fe/H] and [$\alpha$/Fe] abundances.
For the analysis we retained 1160 well-fit spectra with $S/N\geq3$.
Radial velocities are measured to better than $\lesssim1\,\kms$ (median $\sigma_{V_r}=0.2\,\kms$), while typical uncertainties for [Fe/H] and [$\alpha$/Fe] are $0.06$\,dex and $0.04$\,dex, respectively.
Despite the sub-$\kms$ statistical precision, sky-emission lines can show variations of up to $\approx1\,\kms$ across the two camera chips and between different exposures.
Our overall kinematic precision is therefore systematics-dominated at $\approx1\,\kms$ ($\approx0.2\,\rm pix$), comparable to that typically achieved with Hectochelle \citep[e.g.,][]{caldwell2017}.

\section{Stream membership}
\label{sec:membership}

We define a sample of highly probable GD-1 member stars using their \gaia\ proper motions \citep{gdr2}, de-reddened \pans\ photometry \citep{sfd, ps1}, and our measurements of radial velocity and metallicity.
Figure~\ref{fig:members} shows the adopted selection criteria in dashed pink, our spectroscopic sample in light blue, preliminary GD-1 members in medium blue, and final member selection in dark blue, with circles and stars for members in the main stream and the spur, respectively.

Following \citet{pwb}, we start with a generous selection in proper motions (corrected for solar reflex motion): $-10<\mu_{\phi_1}/\masyr<-5.5$ and $-2.5<\mu_{\phi_2}/\masyr<2$ (Figure~\ref{fig:members}, top left).
We further consider stars close to the $\textrm{[Fe/H]}=-2.3$, 12.6\,Gyr isochrone at 8.5\,kpc \citep{choi2016} as more likely GD-1 members (top middle).
The isochrone selection box is tighter around the GD-1's main sequence where the contrast with respect to the Milky Way field is higher, and wider along the red giant branch.
Next, we require GD-1 members to have a small radial velocity offset from the GD-1's orbit\footnote{we derived an updated GD-1 orbit in Section~\ref{sec:kinematics}}, $|\Delta V_r| < 7\,\kms$ (top right).
Finally, we select stars with $-2.37<\feh_{\rm init}<-1.75$ for a final sample of 43 most likely GD-1 stars (Figure~\ref{fig:members}, bottom).
In the interest of producing a pure sample, our spectroscopic selection criteria are rather stringent.
Future analyses may improve the completeness of this sample using probabilistic membership approaches.
The full spectroscopic sample with GD-1 membership flags we developed is publicly available.\footnote{See \url{https://github.com/abonaca/spur_rv}.}

GD-1 stars in the main stream and in the spur have similar radial velocities and chemical abundances, demonstrating that the spur is indeed a part of GD-1.
The stellar population in GD-1 is metal-poor, $\feh=-2.3\pm0.1$ (the initial metallicity, show in Figure~\ref{fig:members}, is slightly higher due to diffusion, $\feh_{\rm init}=-2.1\pm0.2$), and alpha-enhanced, $\afe=0.4\pm0.2$.
Abundance spreads in both \feh\ and \afe\ can be accounted for by the measurement uncertainties in the overall sample, but for a subset of high signal-to-noise stars ($S/N>10$), the variation in \afe\ is two times larger than the measurement uncertainty.
Variations in the abundance of light elements and little \feh\ spread are commonly observed in globular clusters \citep{gratton2019}.
These abundance trends, combined with a low velocity dispersion (Section~\ref{sec:kinematics}), suggest that GD-1 stream is a disrupted globular cluster.

\section{GD-1 kinematics}
\label{sec:kinematics}

\begin{figure*}
\begin{center}
\includegraphics[width=0.9\textwidth]{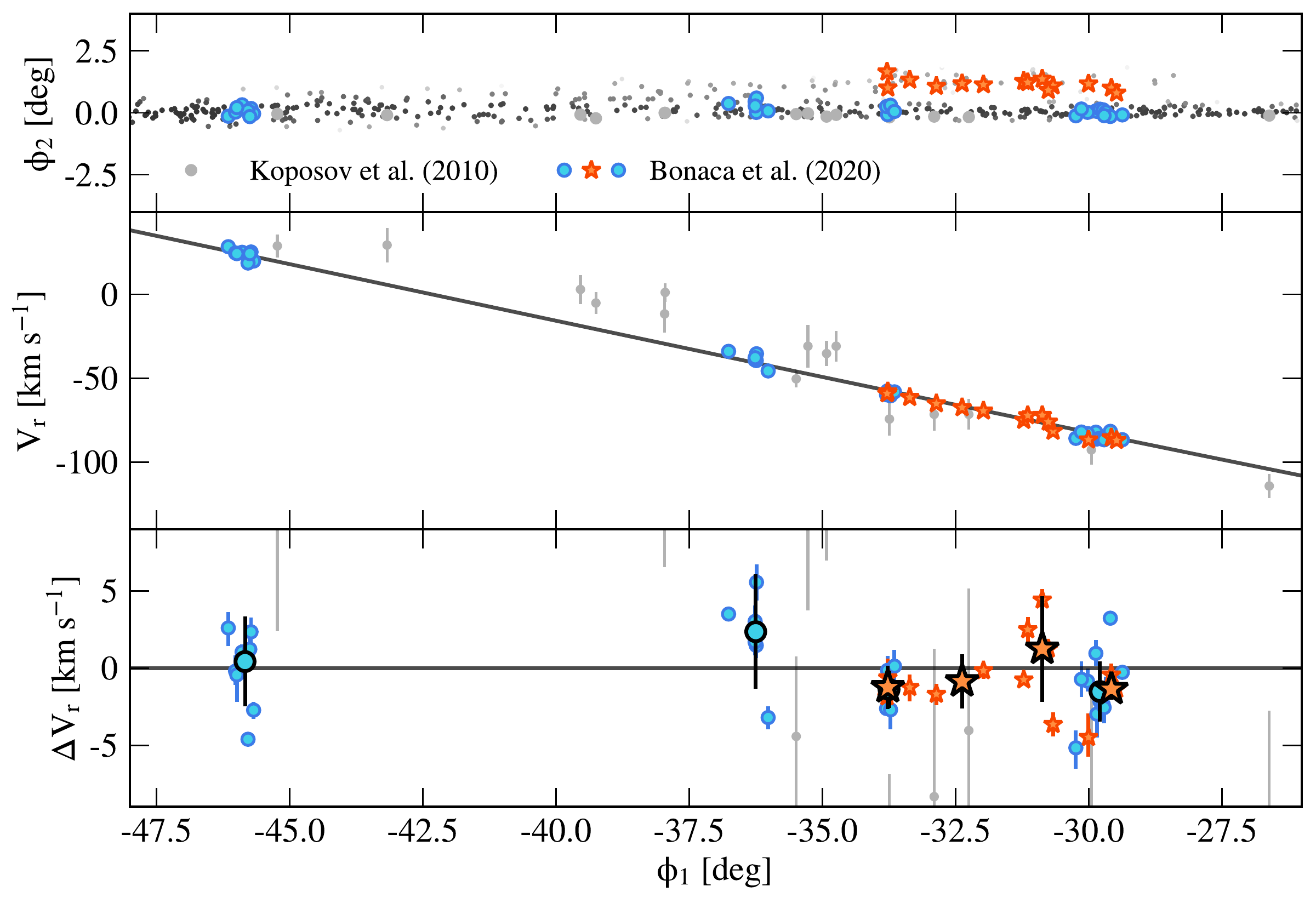}
\end{center}
\caption{Sky positions of spectroscopically identified GD-1 members overplotted on the map of likely stream members (top; gray for \citet{koposov2010} data, blue circles and orange stars for the main stream and the spur data from this paper, respectively).
Both data sets agree that GD-1 has a steep radial velocity gradient (middle), which puts tight constraints on the stream's orbit (black).
At a fixed location along the stream, the median radial velocities of the main stream and the spur are consistent at a level of $\lesssim1\,\kms$ (black-outlined symbols, bottom).
}
\label{fig:vr}
\end{figure*}

We summarize radial velocity structure of the GD-1 stream in Figure~\ref{fig:vr}.
The top panel shows the on-sky distribution of likely GD-1 members identified using \gaia\ proper motions \citep[small points,][]{pwb}, and highlights stars with a measured radial velocity \citep[blue / orange for this work, gray for literature data from][]{koposov2010}.
The second panel shows radial velocity as a function of the $\phi_1$ stream coordinate.
Our data include the first radial velocity measurements in the GD-1 spur (orange stars), and they are consistent with radial velocities in the main GD-1 stream (blue circles).
Our measurements show a strong radial velocity gradient along the stream that is largely consistent with, but somewhat offset from the literature measurements (obtained at a lower-resolution).
We next search for orbits that fit the updated sample of GD-1 radial velocities.

We adopted the GD-1 orbit-fitting procedure from \citet{pwb}, including their fixed Milky Way model similar to \citet{bovy2015} and their compilation of 6D stream data which we augmented with more precise radial velocities from this work.
The radial velocity gradient of the best-fit orbit is shown with a black line in the second panel of Figure~\ref{fig:vr}.
The best-fit orbit has a pericenter at 13.8\,kpc and an apocenter at 21.5\,kpc, making the updated orbital solution slightly more circular, but otherwise similar to the orbit derived in \citet{pwb}.

In the third panel of Figure~\ref{fig:vr} we show the radial velocity offsets from the best-fit GD-1 orbit (black).
Overall, our high-resolution measurements show little deviation from the orbital velocity and reveal a kinematically cold stream with a much lower dispersion than previously measured.
Accounting for measurement uncertainties, the intrinsic velocity dispersion in GD-1 is smaller than $\lesssim1\,\kms$.
Repeat measurements of radial velocities with the same instrumental setup indicate that only slightly higher precision of $0.6\,\kms$ can be achieved with MMT/Hectochelle \citep{cargile2019}, so resolving the velocity dispersion in GD-1 may require higher-resolution spectroscopy.

To quantify the relative motion between the stream and the spur, we compare the median radial velocity of GD-1 members observed in individual Hectochelle fields (large symbols with a black outline, the errorbars are the standard deviation).
At two locations where we observed the main stream and the spur in parallel ($\phi_1=-33.7^\circ, -30^\circ$), the relative radial velocity is smaller than $0.5\,\kms$, comparable to the measurement uncertainty.
To a high degree, the GD-1 spur is comoving with the stream, which puts strong constraints on formation scenarios.

\begin{figure*}
\begin{center}
\includegraphics[width=0.99\textwidth]{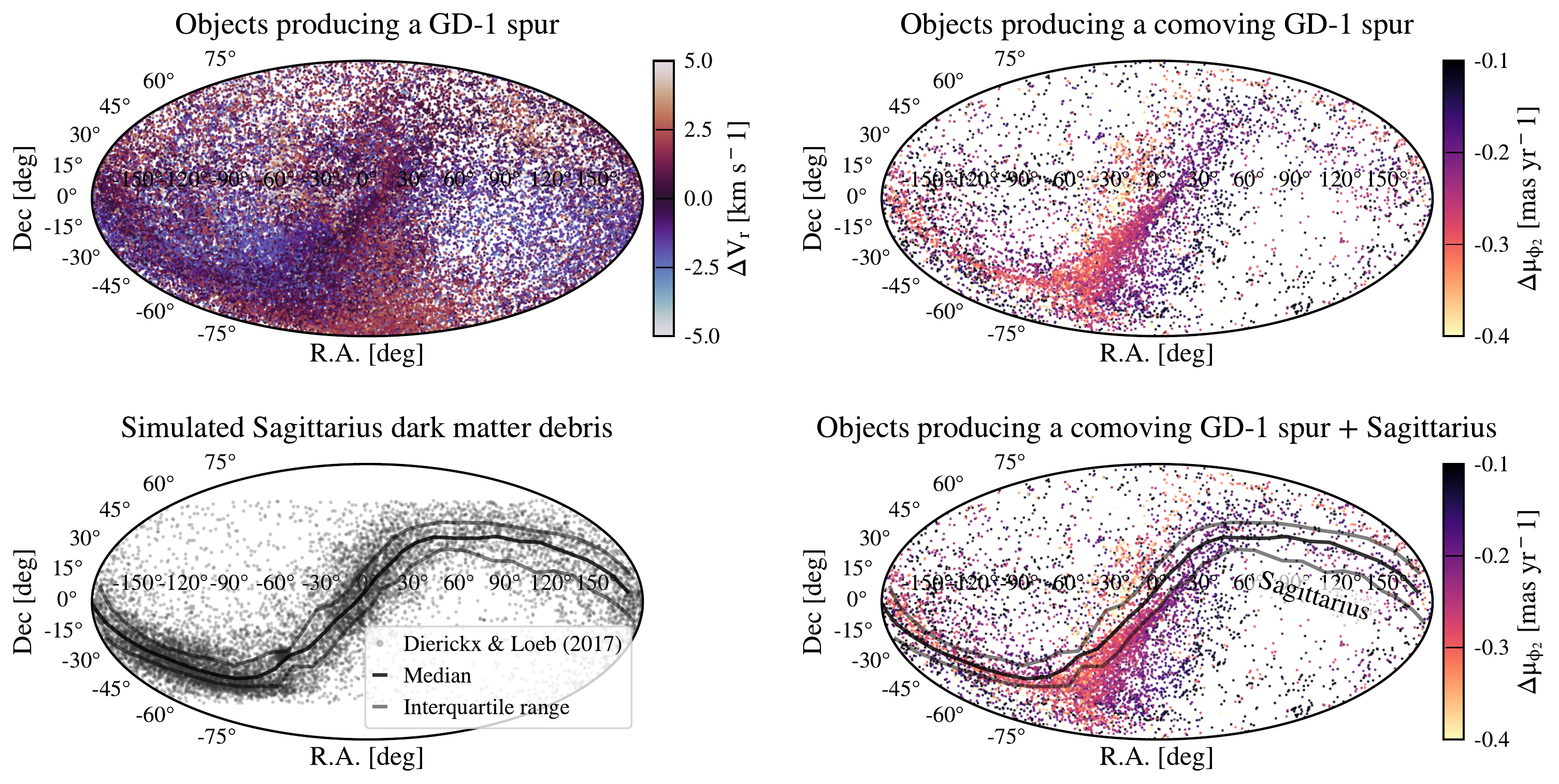}
\end{center}
\caption{Present-day sky positions of objects sampled from a distribution that induces a spur-and-gap morphology after a close encounter with the GD-1 stream, color-coded by the relative radial velocity, $\Delta V_r$, between the stream and the spur at $\phi_1=-33.7^\circ$ (top left, from \citealt{bonaca2019a}).
Solutions that satisfy the measured radial velocity offset, $|\Delta V_r|<1\kms$, are approximately on a great circle (top right), coincident with the distribution of dark matter expected from the disruption of the Sagittarius dwarf galaxy (bottom, gray points from \citealt{dl2017}).
}
\label{fig:skybox}
\end{figure*}

\section{Discussion}
\label{sec:discussion}

We presented high-resolution spectroscopy at eight locations in the GD-1 stellar stream, distributed along the main stream and an adjacent spur.
With the goal of discerning the association between the stream and the spur, we obtained the most precise radial velocities of GD-1 to date (statistical uncertainty $\lesssim0.5\,\kms$).
These data also update the GD-1's orbit (\S\,\ref{sec:kinematics}), and will improve constraints on the Milky Way's gravitational potential in future modeling of GD-1 \citep[e.g.,][]{koposov2010, bowden2015}.
The relative radial velocity between the stream and the spur is small, $\Delta V_r\lesssim1\,\kms$, which, combined with their similar metallicity, $\feh\approx-2.3$, suggests that the spur is a part of GD-1 that has been perturbed from an original orbit along the stream.
We conclude with a discussion of implications that a comoving spur places on its formation mechanism and an outlook for dynamical inferences about the structure of the Milky Way if features like the GD-1 spur are common in other streams.

\citet{bonaca2019a} showed that a stream can develop the spur-and-gap morphology similar to that observed in GD-1 following an encounter with a massive object, and predicted that properties of the encounter can be further constrained with kinematic data.
We revisit perturbed stream models from \citet{bonaca2019a} that were sampled to provide a good quantitative match for the GD-1 morphology, and explore implications of the observed GD-1 kinematics in the context of an encounter scenario.
A spur comoving with the stream prefers models of a closer encounter with a less massive and more compact object than inferred from the stream morphology alone, while the range of allowed impact times and the perturber's total velocity remain similar.
The most substantial improvement that the kinematic data provide is in constraining the perturber's orbit, which determines its present-day location.
In the top left of Figure~\ref{fig:skybox} we show the present-day sky positions of perturber models allowed by stream morphology, color-coded by the relative stream--spur radial velocity, $\Delta V_r(\phi_1=-33.7^\circ)$.
Morphology alone allows for a perturber on a variety of orbits, that result in present-day positions distributed across $\approx30,000\,\deg$.
However, models satisfying a conservative estimate of the relative velocity, $|\Delta V_r|<1\,\kms$, are spatially constrained to $\approx5,000\,\deg^2$ (Figure~\ref{fig:skybox}, top right).

The relative radial velocity measured between the GD-1 stream and its spur improves the localization of the GD-1's perturber by a factor of six, but the resulting area is still too wide for direct follow-up searches.
Better localization is possible if we can measure the radial velocity gradient along the spur, generically expected in interaction models \citep{bonaca2019a}.
Radial velocities we measured in the GD-1 spur show a tentative gradient between $\phi_1=-35^\circ$ and $-31^\circ$ (Figure~\ref{fig:vr}, bottom), however, higher precision is required to fully resolve the gradient.
Further improvements in the sky localization of the pertuber are possible by measuring the relative proper motion between the GD-1 stream and the spur (color-coding in the right panels of Figure~\ref{fig:skybox}).
We expect the precise transverse velocities from \gaia\ (end-of-mission precision $\approx0.25\,\masyr$) or \hst\ (3-year-baseline precision $\approx0.1\,\masyr$) to constrain the perturber's location and enable direct follow-up.

Current localization of the GD-1 perturber suggests it might originate from the Sagittarius dwarf galaxy.
\citet{dl2017} simulated disruption of Sagittarius which results in a distribution of dark matter particles (Figure~\ref{fig:skybox}, bottom left) that spatially overlap with inferred positions of GD-1's perturber (bottom right).
A subset of GD-1 solutions between $\rm R.A.=180^\circ$ and $300^\circ$ further coincide with the expected distances and radial velocities of the Sagittarius debris.
The possibility of GD-1's perturber originating from Sagittarius underlines the importance of accretion and time evolution in the Milky Way halo.
Specifically, future tests of dark matter substructure based on stream gaps will need to account for recently accreted subhalos in addition to the relaxed, isotropic population that has been assumed so far \citep[e.g.,][]{erkal2016, banik2019}.

A globular cluster or a dark-matter subhalo associated with Sagittarius are plausible culprits to produce the spur-and-gap morphology in GD-1, while the Sagittarius dwarf itself is too massive.
Globular clusters appear more likely candidates due to the compact size we infer for the perturber, but none of the known clusters come closer than 1\,kpc to the GD-1 impact site during the past 2\,Gyr (based on the analysis from \citealt{bonaca2019a} with the updated orbit of GD-1 and the 6-dimensional cluster positions from \citealt{baumgardt2019}).
Still, the scenario in which GD-1 was perturbed by a globular cluster needs to be further tested, as the census of globular clusters may be incomplete and the true gravitational potential likely deviates from the idealized model we used so far.
If a luminous perturber is conclusively ruled out after these considerations have been taken into account, a dark-matter subhalo associated with Sagittarius remains a viable perturber, and its inferred high density might signal self-interacting dark matter \citep[e.g.,][]{kahlhoefer2019}.

GD-1 is a stellar stream displaying many surprising features, which has sparked a discussion of additional processes to explain different aspects of the data.
For example, \citet{deboer2019} explored models in which GD-1 is perturbed by the Sagittarius dwarf galaxy.
A strong interaction with Sagittarius can launch a long spur that remains closely aligned with GD-1 before detaching from the main stream (which reproduces the widening of GD-1 at $\phi_1\lesssim-45^\circ$).
On the other hand, \citet{malhan2019c} discovered a low surface-brightness stream, Kshir, that intersects GD-1 at $\phi_1\approx-20^\circ$.
This cross-point is sufficiently close to the spur-and-gap feature that Kshir might have affected their formation.
Alternatively, \citet{webb2019} suggested that the gap at $\phi_1\approx-40^\circ$ may not be a signature of an impact, but rather the location of the GD-1 progenitor's final disruption.
In that case, the spur could be a result of substructure in the progenitor \citep[e.g.,][]{carlberg2018}, instead of forming through an external perturbation.
Overall, such processes impart large velocity kicks, so their role in GD-1's history needs to be reconsidered in the context of a comoving, kinematically cold spur.

We have shown that compact objects can be located in the Milky Way halo by dynamical modeling their impact on cold stellar streams like GD-1.
The prospect of subhalo localization would revolutionize the study of dark matter in the Milky Way.
Instead of inferring the nature of dark matter through the total abundance \citep[e.g.,][]{carlberg2013} or the mass function of dark-matter subhalos \citep[e.g.,][]{banik2019}, multi-wavelength observations of individual subhalo candidates would enable direct tests of different dark matter models \citep[e.g.,][]{daylan2016}, and add dark matter to the domain of multi-messenger astronomy.

\vspace{0.5cm}
It is a pleasure to thank Marion Dierickx for providing Sagittarius models, and Daniel Eisenstein, Doug Finkbeiner, Lars Hernquist, Sean Moran, and Hans-Walter Rix for valuable discussions.
Observations reported here were obtained at the MMT Observatory, a joint facility of the Smithsonian Institution and the University of Arizona.
This project was developed in part at the 2019 Santa Barbara Gaia Sprint, hosted by the Kavli Institute for Theoretical Physics at the University of California, Santa Barbara.
This research was supported in part at KITP by the Heising-Simons Foundation and the National Science Foundation under Grant No. NSF PHY-1748958.

\facility{MMT(Hectochelle)}

\software{
\package{Astropy} \citep{astropy, astropy:2018},
\package{gala} \citep{gala},
\package{IPython} \citep{ipython},
\package{matplotlib} \citep{mpl},
\package{numpy} \citep{numpy},
\package{scipy} \citep{scipy}
}

\bibliographystyle{aasjournal}
\bibliography{spur_rv}

\end{document}